\documentclass[a4paper,fleqn,usenatbib]{mnras}

\usepackage{newtxtext,newtxmath}

\usepackage[T1]{fontenc}
\usepackage{ae,aecompl}
\usepackage{pdflscape}	

\usepackage{graphicx}	
\usepackage{amsmath}	
\newcommand{\lam}{$\,\mu$m}
\newcommand{\hii}{{\sc h~ii}}

\title[New models for PAH emission in galaxies]{
The properties of Polycyclic Aromatic Hydrocarbons in galaxies:
constraints on PAH sizes, charge and radiation fields}

\author[D. Rigopoulou et al.]{
D. Rigopoulou,$^{1}$\thanks{E-mail:dimitra.rigopoulou@physics.ox.ac.uk}
 M. Barale,$^{1,2}$
 D. C. Clary,$^{3}$
 X. Shan,$^{3}$
 A. Alonso-Herrero,$^{4}$
 \newauthor
 I. Garc{\'i}a-Bernete,$^{1}$
L. Hunt,$^{5}$
B. Kerkeni,$^{1,6,7}$
M. Pereira-Santaella,$^{1,8}$
 and P.F. Roche$^{1}$
 \\
$^{1}$Astrophysics, Department of Physics, University of Oxford, Keble
Road, Oxford, OX1 3RH, UK\\
$^{2}$Universit\'e de Rennes, CNRS, ISCR (Institut des Sciences Chimiques de
Rennes) - UMR 6226, F-35000 Rennes, France\\
$^{3}$ Physical \& Theoretical Chemistry Laboratory, University of
  Oxford, South Parks Road, Oxford OX1 3QZ, UK\\
$^{4}$ Centro de Astrobiolog{\'i}a, CSIC-INTA, ESAC Campus, E-28692,
  Villanueva de la Ca\~nada, Madrid, Spain\\
$^{5}$ INAF, Osservatorio Astrofisico di Arcetri, Largo E Fermi 5,
50125 Firenze, Italy\\
$^{6}$ D{\'e}partement de Physique, LPMC Facult{\'e} des Sciences de
Tunis, Universit{\'e} de Tunis el Manar, Tunis 2092, Tunisia\\
$^{7}$ISAMM, Universit{\'e} de La Manouba, La Manouba 2010, Tunisia\\
$^{8}$Centro de Astrobiolog{\'i}a (CSIC-INTA), Ctra. de Ajalvir, km
4, 28850, Torrej\'on de Ardoz, Madrid, Spain
}

\date{Accepted 2021 March 31. Received 2021 March 31; in original form
  2020 July 10}

\pubyear{2020}

\begin{document}
\label{firstpage}
\pagerange{\pageref{firstpage}--\pageref{lastpage}}
\maketitle

\begin{abstract}
Based on theoretical spectra computed using Density Functional Theory
we study the properties of Polycyclic Aromatic Hydrocarbons
(PAH). In
particular using bin-average spectra of PAH molecules with varying
number of carbons we investigate how the intensity of the 
mid-infrared emission bands, 3.3, 6.2, 7.7 and 11.3 $\mu$m, respond
to changes in the number of carbons, charge of the molecule, and the hardness of
the radiation field that impinges the molecule. We confirm
that the 6.2$/$7.7 band ratio is a good predictor for the size of
the PAH molecule (based on the number of carbons present).
We also investigate the
efficacy of the 11.3$/$3.3 ratio to trace the size of PAH molecules
and note the dependence of this ratio on the hardness of the radiation
field. While the ratio can potentially also be used to trace PAH
molecular size, a better understanding of the impact of the underlying
radiation field on the 3.3 $\mu$m feature and the effect of the
extinction on the ratio should be evaluated.

The newly developed diagnostics are compared to band ratios measured in a
variety of galaxies observed with the Infrared Spectrograph on board the
{\it Spitzer Space Telescope}. We demonstrate that the band ratios can be used to
probe the conditions of the {\it interstellar medium} in galaxies and
differentiate between environments encountered in
normal star forming galaxies and Active Galactic Nuclei. Our work
highlights the immense potential that PAH observations with the {\it James Webb
  Space Telescope} will have on our understanding of the PAH emission
itself and of the physical
conditions in galaxies near and far.

\end{abstract}




\section{Introduction}

Strong emission features at 3.3, 6.2, 7.7, 8.6, 11.3 and 12.7 {\lam}
are commonly attributed to mixtures of Polycyclic Aromatic
Hydrocarbons (PAHs) which are large molecules resulting from fusion of
aromatic rings and related species \citep{tiel08}. Upon absorption of
ultraviolet (UV) photons, PAH molecules become highly vibrationally
excited and subsequently relax through emission of infrared (IR)
photons at specific wavelengths \citep{dl07}. The
absorption efficiency of PAHs has been modeled based on astrophysical
observations, laboratory measurements, and quantum theory 
(e.g., \citealp{job92, vers01, dl01, dl07, mall07}). In addition to being major radiative
coolants of the Interstellar medium (ISM), PAHs are also responsible
for most of the photoelectric heating of the gas in photodissociation
regions (PDRs) and the neutral ISM, due to their high cumulative
surface area \citep{baktil94, holtiel97}.
As such, PAHs play an important role in the energy balance of
the ISM.

The ubiquity of these mid-infrared (mid-IR) features in a wide variety
of astrophysical objects and enrivonments provides a useful diagnostic
of the physical conditions in these sources.  Mid-IR spectroscopic
observations with ESA's Infrared Space Observatory \citep{kess96}
and NASA's Spitzer Space Telescope \citep{wern04} of a
number of galactic and extragalactic sources have confirmed that PAH
features are found to dominate the spectra of evolved stars
\citep{blom05} all the way to spectra of entire galaxies
\citep{brandl06, smith07}.  From an extragalactic point of view, the
strengths of the 7.7 and 6.2 $\mu$m features have been proposed as
tracers of star-formation (e.g., \citealp{rigo99, peet04}).
The diagnostic power of these features to trace star
formation has been thoroughly investigated (e.g. \citep{shipl16}) and applied
out to the highest redshift sources (e.g., \citealp{huang07, pope13, riech14}).
However, there are claims that the tracers are affected
by global galaxy parameters such as the metallicity
\citep{engel08, hunt10}.

The picture becomes more complicated for galaxies that host an Active
Galactic Nucleus (AGN). Galaxies with AGN tend to have low 6.2 and
11.3 $\mu$m
PAH equivalent width (EQW) (e.g., \citealp{roche89, rigo99, tran01, sturm00,
  desai07, ds10}) due to the presence of a significant hot
dust continuum and also because the hard AGN photons may destroy the small
PAH molecules. However, more recent ground-based observations detect
the 11.3 {\lam} PAH feature at distances of hundreds to tens of
parsecs from the AGN
(e.g., \citealp{honig10, gonzmar13, aah14, aah16}).
Moreover, there are indications that the AGN may
also be the source of PAH excitation \citep{jens17}.
A recent review by \citet{li20} highlights our current
understanding of the impact of specific galaxy properties on
the strength of the PAH bands.

The detailed characteristics of PAHs, such as their central wavelength,
shape and the intensity ratio between different bands, are known
to vary \citep{peet04}.  These variations primarily reflect
changes in the structure of the PAH molecules in response to 
diverse astrophysical environments. Because each PAH band is
attributed to a specific vibrational mode, the ratio between different
bands could be used as a diagnostic of PAH properties.
For instance, it is well known that the 3.3 $\mu$m
PAH feature is due to the radiative relaxation of C-H stretching
modes, the 6-9 $\mu$m features originate from C-C stretching modes
while the 11.3 $\mu$m feature originates in the C-H out-of-plane
bending modes. Experimentally, it has also been confirmed
that the C-C modes are intrinsically weak in neutral PAHs compared to
ionised PAHs \citep{allam99, bausch02, kimsay02}.
Therefore the 6-9 $\mu$m features will be much
more prominent for ionised PAHs, while the converse is true for the
3.3 and 11.3 $\mu$m bands. Consequently, the ratios between the C-C
and the C-H feature intensities depend on the charge of the PAHs,
which is directly related to the physical conditions in the environment
where the emission is originating.

A number of authors have presented evidence for the existence of 
variations between the strength of PAH
bands in different astrophysical environments but also within a given
source \citep{hony01, gall08, peet17}.
\citet{bregtem05} studied the variation
of the 7.7$/$11.3 ratio in reflection nebulae and found a correlation of 
this band ratio with the ratio G$_{\rm 0}$/n$_{\rm e}$ between the integrated
intensity of the UV field, G$_{\rm 0}$, and the electron density,
n$_{\rm e}$. However, \citet{smith07} concluded that this ratio is
relatively constant among pure starburst galaxies but varies by a
factor of 5 among galaxies having a weak AGN. 
They interpret this effect as a selective destruction of the smallest
PAHs by the hard radiation arising from the accretion disk, ruling out
an explanation in terms of ionization of the molecules, in these
particular environments. 

The use of PAH band ratios as
diagnostics of the physical conditions of the ISM has been established
in previous works (e.g., \citealp{hony01, sloan07, gall08, stopeet17}). 
In this paper we use theoretically computed PAH spectra
to investigate how the PAH molecular size (in this work parameterized 
by the number of carbons N$_{\rm c}$), the PAH charge (ionised
vs. neutral) and the average photon energy affect the 
relative intensities of each PAH band.
The aim of this work is to
assess through our analysis the use of PAH band ratios 
as diagnostics of the physical conditions of the ISM in galaxies.
In particular we wish to investigate how the measured PAH band ratios
can be used to infer
PAH molecular sizes, charge and the hardness of the
radiation field that excites the PAHs.
As we are using PAH band ratios measured in relatively large regions
($\sim$kpc scale) in  galaxies (using
spectra from ISO and Spitzer) the focus
of this work is not to {\it identify specific PAH or even classes
of PAH molecules} but instead to
carry out a qualitative investigation of the PAH band ratios.

The
paper is arranged as follows.
In Section 2 we discuss the theoretical
approach and the methodology followed to compute PAH theoretical spectra 
using Density Functional Theory (DFT) as well as the emission model
used. A detailed descrription of the molecules included in this work
is also presented.
In Section 3 we present the model PAH spectra and measure the intensities of
the various features. Sections 4 and 5  
contain the results and discussion of their astrophysical implications
followed by conclusions which are summarised in Section 6.

\section{Methodology}

When an interstellar PAH molecule absorbs a UV photon the energy is quickly
transferred to the vibrational levels and the molecule proceeds to an
excited state. Once excited, the molecule will
cool by fluoresence and emit IR photons at the frequencies that
correspond to specific vibrational modes \citep{allam99, baktil94}.
It has been shown \citep{allam89}
that the IR fluoresence of such molecular sized species
is well approximated by the product of the
absorption cross
section convolved with the Planck function as was proposed by
\citet{leger84}.
This approach has formed the cornerstone of most
theoretical modeling studies of interstellar PAHs. 
Since some vibrational modes (in particular stretching
of C -- C and bending of C -- H bonds) are common to most PAHs and also occur
at similar frequencies, a significant population of PAHs will emit very strongly
at those common frequencies. As a consequence, the observed emission
bands are overall fairly similar in appearance in different lines of
sight.

The intrinsic spectral characteristics of PAHs, such as their
vibrational frequencies, can be computed with a variety of techniques
summarised in \citet{tiel08}.
The focus of these methods is to solve Schr{\"o}dinger's
equation for molecular systems with many electrons using a specific
set of approximations.  All of the theoretical PAH spectra used in this work
have been computed using
DFT. The method provides a cost-effective
approximate solution of Schr{\"o}dinger's equation and has made very
useful predictions of spectra of
PAHs which compare favourably with available experimental data \cite{tiel08}. 

\subsection{Selection Criteria}

A large number of DFT calculations of PAH vibrational frequencies have
been reported in the literature.  The  molecules used in the present study
are built with the Gaussview 5 software and 
the DFT calculation were performed using Gaussian 09 \citet{fri09} 
together with the B3LYP functional along with the 4-31G
basis set. The functional enables 
the optimisation of the molecular structure while the basis set
provides the mathematical functions that describe the molecular
orbitals. These are expressed as linear combinations of the
basis whose size controls the accuracy of the solution and the
computational cost. A detailed discussion of the various methods
used to compute PAH virbational frequencies and their characteristics
is included in a forthcoming paper (Kerkeni et al, in prep).
Harmonic frequencies and IR intensities have
been computed for large solo-containing neutral and cationic forms of
PAH molecules.
For the present study we have also used spectra from
the NASA Ames PAH IR Spectroscopic Database
\citep{boersma14, bausch18, matt20}. 
The study includes PAH molecules based on the following criteria:\\
1) Number of carbons 20$<$N$_{c}<$400\\
2) Availability of neutral and singly cationic spectra\\
3) Absence of O, N, Fe, Mg in the molecules\\


The entire sample used in this study is
listed in Appendix A.



\subsection{Assumptions and emission model parameters}
The DFT computed transition
frequencies must be convolved with a specific band shape,
line width, and emission temperature to convert them into an
emission spectrum similar to those observed in astrophysical
environments. 
In
what follows
we briefly discuss how the DFT results for frequencies
and intensities can be converted into synthetic models for the PAH emission
occurring in astrophysical environments.

{\it Band profiles and band shifts}. 
While profiles from single vibrational transitions show a
Lorentzian shape, the observed bands result from a superposition of
Lorentzians of slightly different peaks and widths. This happens
because of the emission process and of the different temperatures that
each PAH attains upon absorption of a photon.
Observationally,
PAHs have been fit by Drude profiles \citep{smith07}
as well as Lorentzian ones and \citet{peet17}
discussed the issue in more detail.
For the present
study we have adopted Lorentzian profiles for each PAH
transition. 
The adopted profiles can be parameterised as:
\begin{equation}
  \Pi(\rm \nu) = \frac{1/2  \Gamma}{\pi (\nu - \nu_{i})^2 + (1/2 \Gamma)^2}
\end{equation}

where $\Gamma$ (cm$^{-1}$) is the FWHM of the feature and $\nu_{i}$ is the
frequency of the mode i.
The FWHM of the profiles observed in astronomical objects
are found to vary
substantially amongst the different wavelength
regions. \citet{peet04} found that the mid-IR PAH features show
line widths of 10-30 cm$^{-1}$ while for the longer wavelenth bands
(beyond 15 $\mu$m) the features are typically 4 to 8 cm$^{-1}$ \citep{moutou98}. 
Since in this work we focus on the 2.5-15 $\mu$m
region we adopt a FWHM value of 30 cm$^{-1}$ for all the features.

The vibrational frequencies are computed using the harmonic
approximation. To account for limitations in theory and also for
anharmonic corrections and to ensure a better agreement with 
experimental spectra the frequencies {\it need to be scaled}
appropriately.
Typically, {\it a scaling factor} of 0.958 for
B3LYP$/$4--31G is used for simplicity for the entire spectrum, based
on comparison with matrix-isolation data. However, it is known that
this technique  may introduce unpredictable shifts in the band positions
\citep{lang94}. For the spectra used here we have applied a wavelength-specific
scaling factor as follows: 0.956 for wavelengths less than 2 $\mu$m, 0.952 for
wavelengths between 4 and 9 $\mu$m and 0.96 for wavelengths beyond 9
$\mu$m. 

{\it Emission model \& band intensities.}
Relative emission band intensities
depend on the excitation energy (temperature) that the PAH attains and must 
be taken into account when considering the mid-IR
emission spectrum. Several approaches exist (e.g., \citealp{allam89})
the simplest approximation
amounts to multiplying the intensity in each band with the Planck function
at a specific emission temperature:
\begin{equation}
 B (\nu_{\rm i}, T) = \frac {2hc\nu_{\rm i}^{3}} {exp(\frac{hc\nu_{\rm
       i}}{k_{\rm B}T})
   -1}
\end{equation}

where $\nu_{\rm i}$ (cm$^{-1}$) is the frequency of the transition i, T (K)
is the average emission temperature, h (erg s) is Planck's
constant, c (cm s$^{-1}$)
is the speed of light and k (in erg K$^{-1}$) is Boltzmann's constant.
In reality however, the temperature that each PAH molecule will attain
upon absorption of a photon 
depends on a number of parameters including the
energy of the absorbed photon, the 
heat capacity (C$_{\small V}$) and the photo-absorption cross
section ($\sigma_{\rm i}$) of the given PAH.

\citet{legdefour89}
proposed that the IR cooling of PAHs can be described with the
thermal approximation. In this case the PAH molecule can be considered as
a heat bath with an average molecular energy U and temperature
T. Following absorption of a UV photon, the molecule has an
internal energy U(T)
which can be written as:

\begin{equation}
  U(T) = \sum_{\rm i}^{n}  \frac {hc \nu_{i}} {exp (hc\nu_{i} / k_{\rm B} T)-1}
\end{equation}

Upon absorption of a photon with energy hc${\rm \nu}$, the PAH will
attain a peak temperature T$_{\rm p}$ according to:
\begin{equation}
  C_{\rm V} (T) dT = hc {\rm\nu}
\end{equation}

where C$_{\rm V}$(T) is the heat capacity of the PAH molecule which is
related to its internal energy via:

\begin{equation}
  \int\displaylimits_{2.73}^{T_{\rm p}} C_{\rm V} (T) dT \equiv dU(T)/DT
  \end{equation}

where 2.73 is the cosmic microwave background temperature.
 Once the molecule has attained its peak temperature it will cool
 via its various vibrational modes,
 through a so-called radiative cascade. The fractional energy emitted
 in a given mode corresponds to a fall in internal energy
 $\delta$U:

 \begin{equation}
   \delta E_{\rm i} \propto \delta U (T)
 \end{equation}

 The total energy emitted in a single band is obtained by integrating over the
 temperature range from 2.73 K  to the peak temperature (T$_{\rm p}$),
 weighted by the rate of photon absorption given by:
 \begin{equation}
  \sigma_{\rm i} = \int\displaylimits_{\rm 2.73}^{T_{\rm p}}
  \frac{B_{\rm {\nu, i}}^{T}
    \sigma_{\rm {\nu, i}}}{h\nu}
  d\nu
\end{equation}

where $\nu$ is the frequency of the absorbed photon, $\sigma_{\nu}$ is the
frequency-dependent photoabsorption cross-section,
B$_{\rm {\nu, i}}^{T}$ is the Planck function at frequency $\nu$  and
  temperature T and T$_{\rm max}$
represents the high-energy cut-off in the radiation field. For each
molecule, $\sigma_{\nu}$ was taken from \citet{dl01} and
\citet{matt05}.  
  
Under this model each PAH molecule
reaches a different maximum temperature that depends on its heat
capacity, with the subsequent radiative relaxation from that excitation
level or temperature taken into account. The emission model
makes use of the thermal approximation discussed in \citet{vers01}.
\begin{figure}
\centering
  \includegraphics[angle=-90,width=9cm]{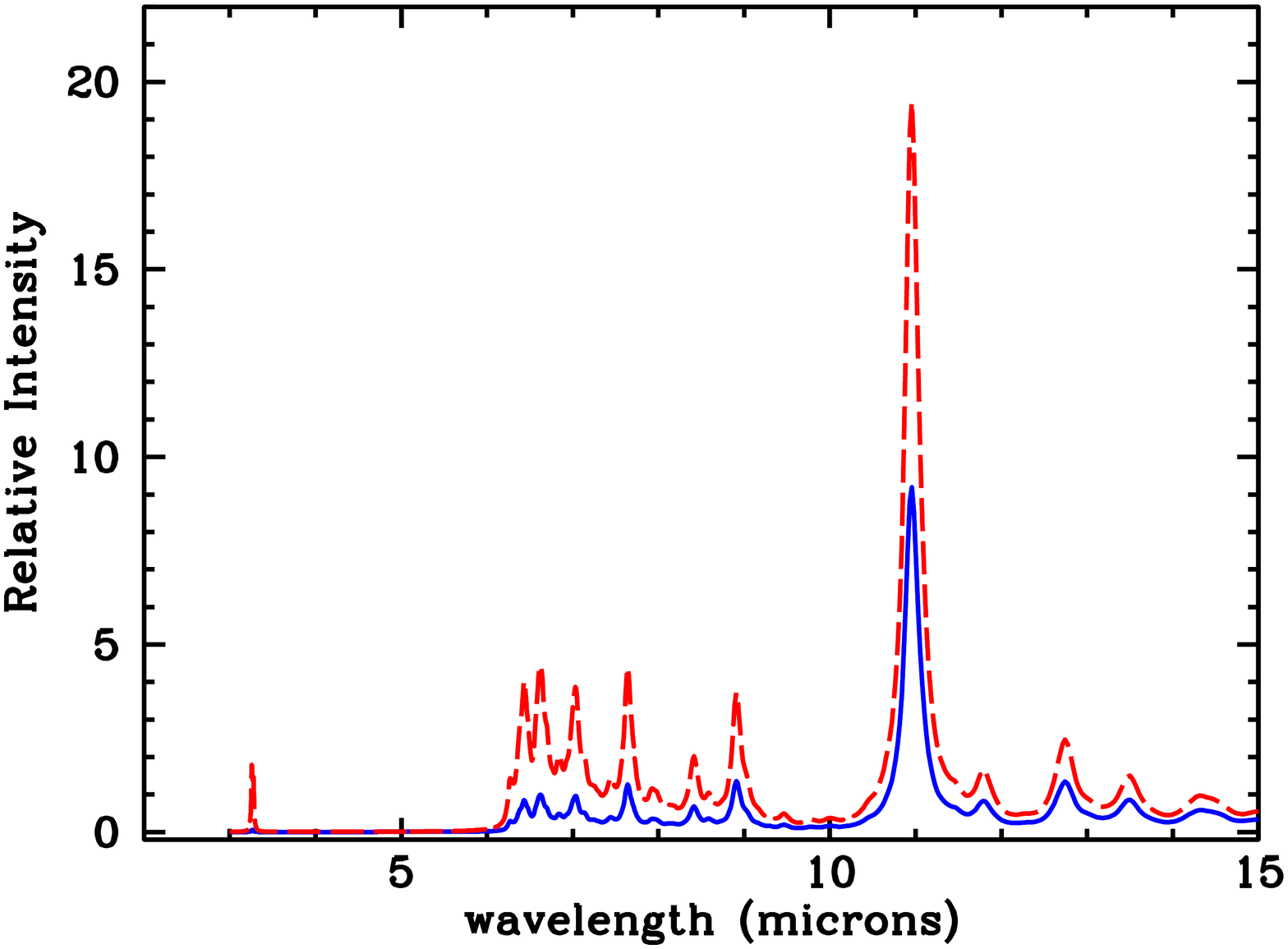}
    \caption{Comparison of the emission features of a PAH molecule
      with N$_{C}$ = 210 which is exposed to a radiation field of 5
      eV (blue line) and 12 eV (red dashed line). The y-axis represents the
      normalised intensities of the features.}
\end{figure}

Clearly, a modification of the temperature distribution of the PAHs
will affect the relative strength of the emission features. One way to
change the PAH temperature is by changing the energy of the UV photons
that are absorbed by PAHs. In an astrophysical setting this would correspond to
e.g. a change in the spectral hardness of the underlying radiation field.
Habing (1968) introduced the dimensionless parameter
\begin{equation}
  \chi = \frac {\nu u_{\rm \nu}} {4 \times 10 ^{-14} erg\, cm ^{-3}}
  \end{equation}

  and determined that
  $\nu u_{\rm \nu} \sim$ 4 $\times$ 10$^{-14}$ erg cm$^{-3}$
  at $\lambda$ =1000 \.A, corresponding to an average photon energy
  of $\sim$12.4 eV for the interstellar radiation field (ISRF) in the
  solar neighbourhood. We
  have used the emission model outlined above to generate
  PAH spectra corresponding to (absorbed) photons at various
  energies.
  In Fig. 1 we show an example of the
  resulting PAH spectra corresponding to a neutral large PAH molecule
  (with N$_{C} =$ 210 carbon atoms) when it
  absorbs a photon of 5 eV (blue line) and one of a 12.4 eV (red
  dashed line).
  As is evident from Fig. 1 the peak intensity of each emission
  feature varies depending on the energy of the photon that was
  absorbed. For example, the peak intensity of the 3.3 $\mu$m feature
  varies by a factor of (I$_{3.3(12.4 eV)}/$I$_{3.3(5 eV)}\sim$11,
  that of the
  11.3 $\mu$m feature varies by a factor of
  (I$_{11.3(12.4 eV)}/$ I$_{11.3(5 eV)} \sim$2 while the ratio of the
  intensities of the 6.2 and 7.7
  $\mu$m features is less sensitive to changes in the photon energy.
  

\section{The PAH spectra}

Determining the nature, composition and exact make-up of the emitting PAH
population in astronomical objects is far from trivial. While spectral
decomposition has been demonstrated as a promising tool for identifying
classes of PAHs in galactic sources (e.g., \citealp{boers13, bausch18}) this method
is not suitable for galaxies given that the beam-averaged
emission encompasses a wide range of PAH molecules of varying size, charge
and composition. In this work we adopt a slightly different approach: by
examining the theoretically computed PAH IR spectra of different size,
geometry, charge and emission model, our aim is to determine how the
different PAH emission
bands change as we step through changes in the properties of the PAH molecules.
For the work presented here
the focus is not on identifying specific species but establishing how
the properties of PAH molecules influence the
strength of each of the most prominent emission bands and how external
conditions such as the strength of the radiation field to which a PAH
molecule may be exposed affects the strength of these bands.
For this purpose, we have ensured that, all the
DFT spectra used in the analysis show broadly the
characteristics that dominate the observed mid-IR spectra of galaxies,
namely, 
the presence of a wide underlying plateau in the 6-9 $\mu$m 
and the strong 11.3 $\mu$m PAH expected to arise mainly from large
neutral molecules.
\begin{figure}
\centering
  \includegraphics[width=9cm]{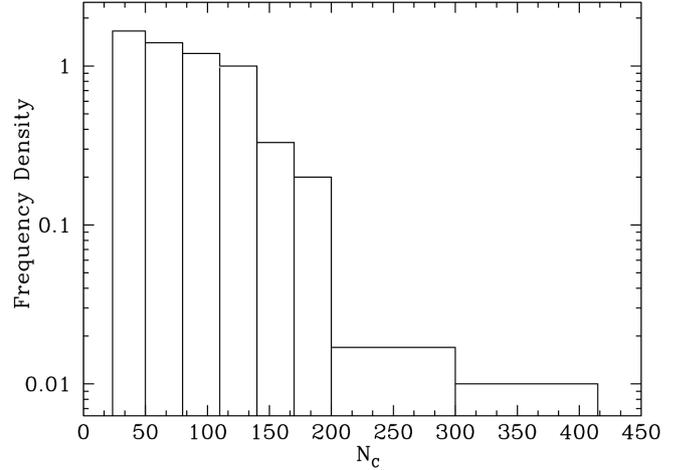}
    \caption{ Histogram depicting the intrinsic size distribution of
      all neutral PAH molecules in this study. Species have been
      placed into bins of unequal width size as discussed in the
      text.}
  \end{figure}

A histogram of the
distribution of all PAH molecules used in
this work, as a function of size
(parameterised by the number of carbons N$_{\rm C}$), is shown in
Fig. 2. Since more PAH molecules with N$_{\rm C}<$150 are available in
our PAH sample
(compared to larger molecules)  the bins in the histogram have unequal widths.
Because of that the vertical axis encodes the
frequency density per unit bin size, which is defined as the ratio of
frequency over width for
each bin \citep{ven02}.
The plot illustrates that the number of large PAH molecules drops
significantly beyond N$_{C}>$200 and we discuss this issue further in
Section 5. 

The computed spectra were all processed following the steps
outlined in Section 2. The emission model discussed in Section 2 was
applied whereby the PAHs 
are excited upon absorption of photons of a certain energy. In the present work we
considered emission models corresponding to photons of a range of
energies from ISRF to 10$^{3} \times$ ISRF.
We then measured the 3.3, 6.2, 7.7, 8.6 and 11.3 $\mu$m
emission band intensities for all the PAH spectra. 
The 
emission band intensities were measured using the publicly available software
CASSIS (cassis.irap.omp.eu) in much the same way that we measure
observed spectra from astronomical sources.
\begin{figure}
\centering
  \includegraphics[width=9cm]{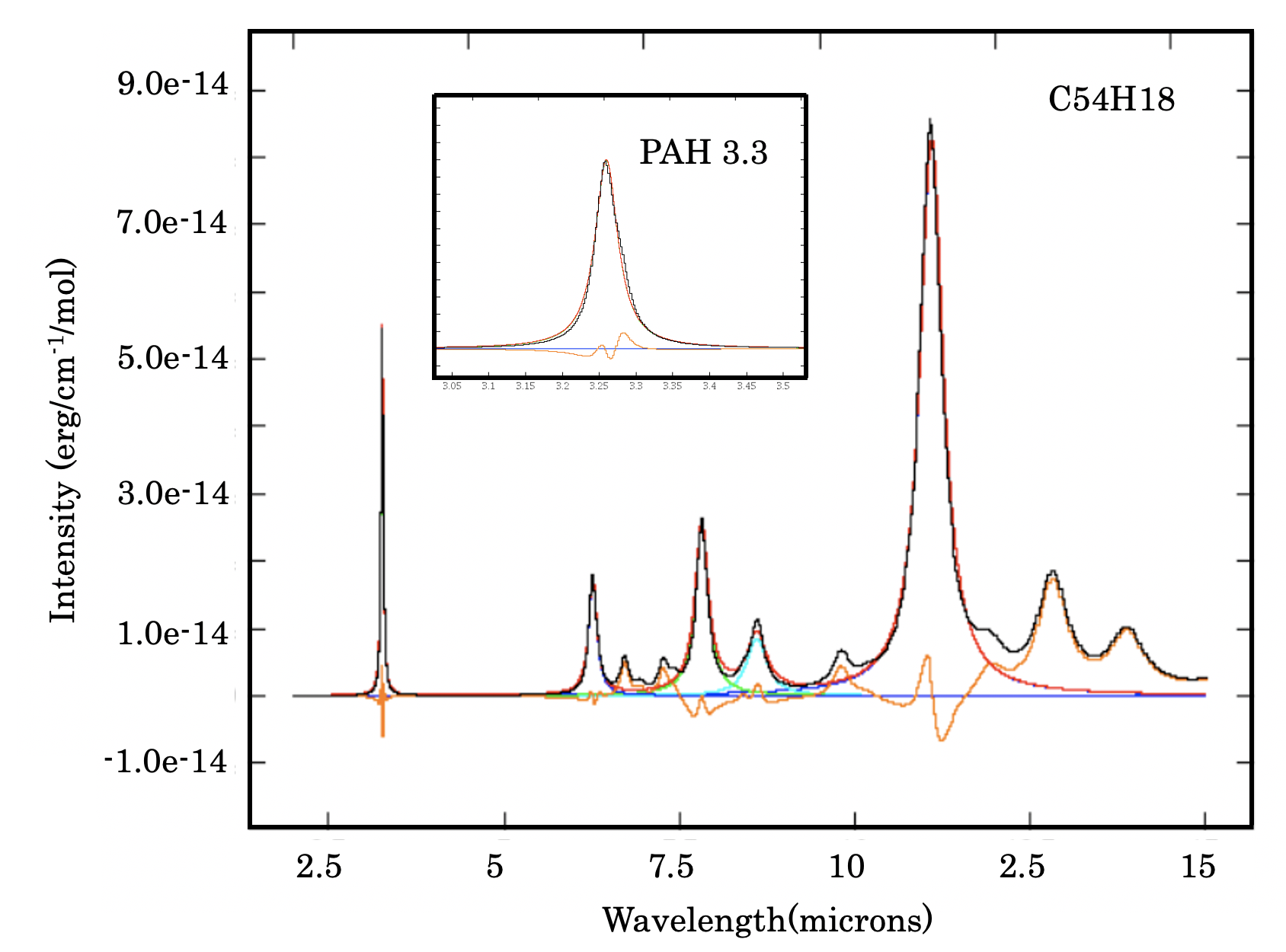}
    \caption{An example of the fitting procedure adopted to measure
      the intensity of the bands of a DFT-computed  PAH spectrum
      (black line) with N$_{C}$=54 (C$_{54}$H$_{18}$).
      exposed to the ISRF hardness. Each spectral band is fitted
      assuming a Lorenzian (corresponding to different colours). The
      inset figure shows the fit for the 3.3 $\mu$m band. The
      orange line denotes the residuals of the fit. The
      width, height and central wavelength are left as free parameters
    during the fit.}
\end{figure}

Astronomical mid-IR PAH features are found on top of broad underlying
plateaus, which represent continuum emission from small grains heated by
stellar light. Different methods have been used to separate PAH
features from the underlying warm dust continuum either by fitting a
spline through fixed points \citep{peet04} or by fitting
Drude profiles (e.g. PAHFIT, \citealp{smith07}) or Lorenztian
profiles \citep{boula98}. However, as \citet{gall08}
pointed out although different methods lead to different implied
band intensities, PAH intensity band ratios
should be largely insensitive to the specific method. 
For the present study we treat the DFT spectra as ``real'' astronomical
spectra. We use a $1^{st}$ order
polynomial to account for the underlying continuum. 
We then identify the wavelength at which each feature peaks and use
this as a first guess for the line fits.
A Lorentzian function is then fit to each PAH emission feature which
is  centered at the
wavelength identified for each feature. The width and height of the Lorentzian
were left as free parameters during
the fit. Fig. 3 shows an example of the fitting procedure for C$_{54}$H$_{18}$. Since
DFT-computed PAH spectra exhibit shifts in the positions of the main
features, depending on the presence of even$/$odd carbons \citep{ricca18}
as well as the geometry, we do not fix the central
wavelength of the features, but rather use it as a first guess for the fit.
This is particularly important
in the location of the 11.3 $\mu$m feature (which is made up of
sub-features in the entire 11.0 -11.3 $\mu$m range). Hence, particular
care was taken during the fits so that the integration was done under
the appropriate range
and for this
reason the central wavelength
was left as a free parameter during the fits.
Our method measures
intensities in the 
DFT-computed spectra in a similar way to that followed for observed
spectra,
with the caveats outlined above.  Our method for
measuring band intensities differs from that followed by \citet{mara20} where the
emission band fluxes are determined by summing over the flux between
pre-determined wavelength ranges.
\begin{figure*}
 \centering
 \includegraphics[width=17cm, angle=0]{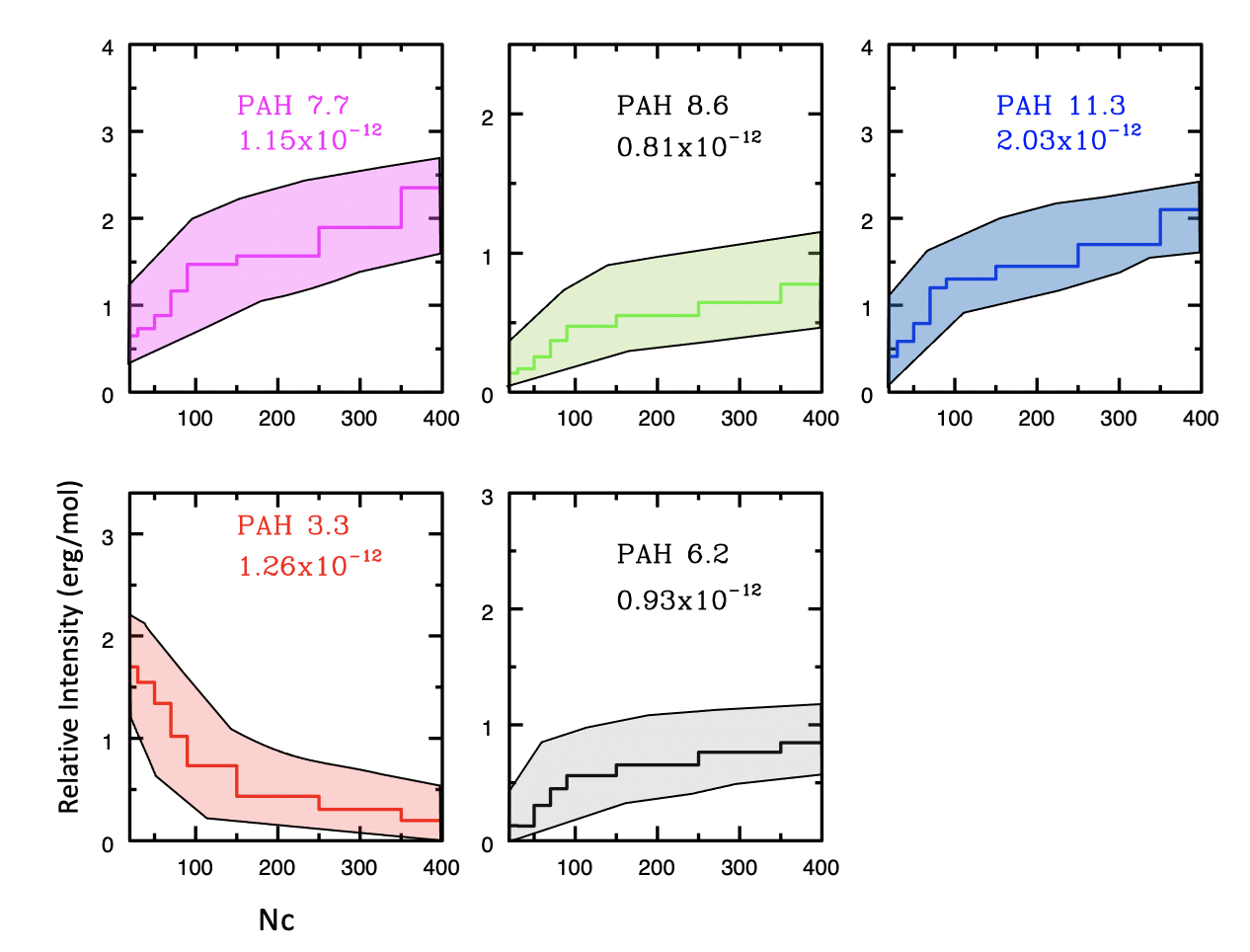}
    \caption{Intensities for neutral PAHs (3.3, 6.2, 7.7, 8.6 and
      11.3 $\mu$m) as a function  of N$_{\rm C}$, with normalisations shown
    in each panel.
      The average value in each bin
     is denoted with the line while the shaded regions denote
     the  spread of
     values in each bin. PAHs with  intensity values greater than 3$\sigma$ have
     been excluded.
     The intensities have been calculated for the ISRF
      radiation field.  Colour coding:
      3.3$\mu$m (red), 6.2 $\mu$m (grey), 7.7$\mu$m (magenta), 8.6 $\mu$m
      (green), 11.3 $\mu$m (blue). }
  \end{figure*}


For each of the individual bins shown in Figure 2 we then calculate a
 composite PAH spectrum. This was done by averaging the spectra in
 that bin and 
that means for every
bin (and every PAH feature) we determined the arithmetic mean by
dividing the average value by the number of molecules that were
assigned to that bin.
If for a specific PAH
molecule its flux value was found to deviate by more than 3$\sigma$ from
the average value determined for the feature then
the spectrum from that PAH molecule was discarded from that bin.
Therefore, each bin is represented by the averaged spectrum of all the
molecules in that bin.

Fig. 4 shows the relative intensities of the 
3.3, 6.2, 7.7, 8.6, 11.3 $\mu$m PAH emission bands calculated
based on the ISRF emission model. The Figure shows the average
intensities as well as the 3$\sigma$ spread around this value for
every bin (see Fig. B1 for the same plot for cations).
All five emission bands show a dependence on N$_{C}$ although with noticeable
scatter which is more pronounced for bands 6.2 and 7.7 $\mu$m. The
least scatter appears in the 11.3 $\mu$m band which is perhaps not
surprising given that neutral PAHs contribute significantly to the
emission in the band.
The intensity of the 3.3 $\mu$m band decreases
with increasing number of carbons N$_{C}$ as expected since the 3.3
$\mu$m band is most sensitive to the smallest PAHs \citep{schutte93}.


\section{PAH Band Ratios: size, charge, hardness of the radiation field
}

The spectral bands and their variations can be used to
probe the size, charge, composition and the underlying radiation field
of PAHs. In particular, the size distribution of PAH molecules is important with
large molecules contributing more to bands at longer wavelengths
whereas smaller PAHs are responsible for the emission at shorter
wavelengths \citep{allam89}.

\subsection{Band ratios vs PAH size distribution}

We start off by examining the dependency of the various intensity band
ratios on the size of the PAH molecules.  The 6.2 $\mu$m and 7.7 $\mu$m bands
are attributed to stretching modes of C -- C bonds, while the 3.3, 8.6
and 11.3 $\mu$m bands are due to
the stretching modes, in-plane
bending modes, and out-of-plane bending modes of C -- H bonds,
\citep{allam89}. Hence, the
ratios of different PAH bands arising from the same vibrational modes
(e.g. 6.2$/$7.7 and 11.3$/$3.3)
could be used to infer the size distribution of PAHs \citep{jdm90, sales10}.
A number of studies aimed at determining PAH size distribution 
in the Spitzer era have relied upon the use of the 6.2 and 7.7 $\mu$m
features as these are the brightest ionic features in the mid-IR spectra
of galactic and extragalactic sources. \citet{dl01} introduced the 6.2$/$7.7 and 11.3$/$7.7 ratios as indicators of
PAH size and ionization, respectively. The \citet{dl01} models are based on
numerical calculations using realistic estimates of  absorption cross sections 
depending on the number of C atoms, the H$/$C ratio, and
the ionization state of the PAHs. Absorption cross sections have
been taken from \citet{allam99}.
The \citet{dl01} analysis
follows the creation of a model PAH grain of a given size which is
subsequently exposed to radiation
fields of varying strengths followed by calculation of the resulting
emission spectrum. 
Fig. 5 shows the intensity ratio of 6.2$/$7.7 as a function of
N$_{C}$ for the DFT computed PAH molecules studied here.
Each point in the plot represents the PAH band ratio measured
  from the averaged spectrum for that bin.

\begin{figure}
\centering
\includegraphics[width=9cm]{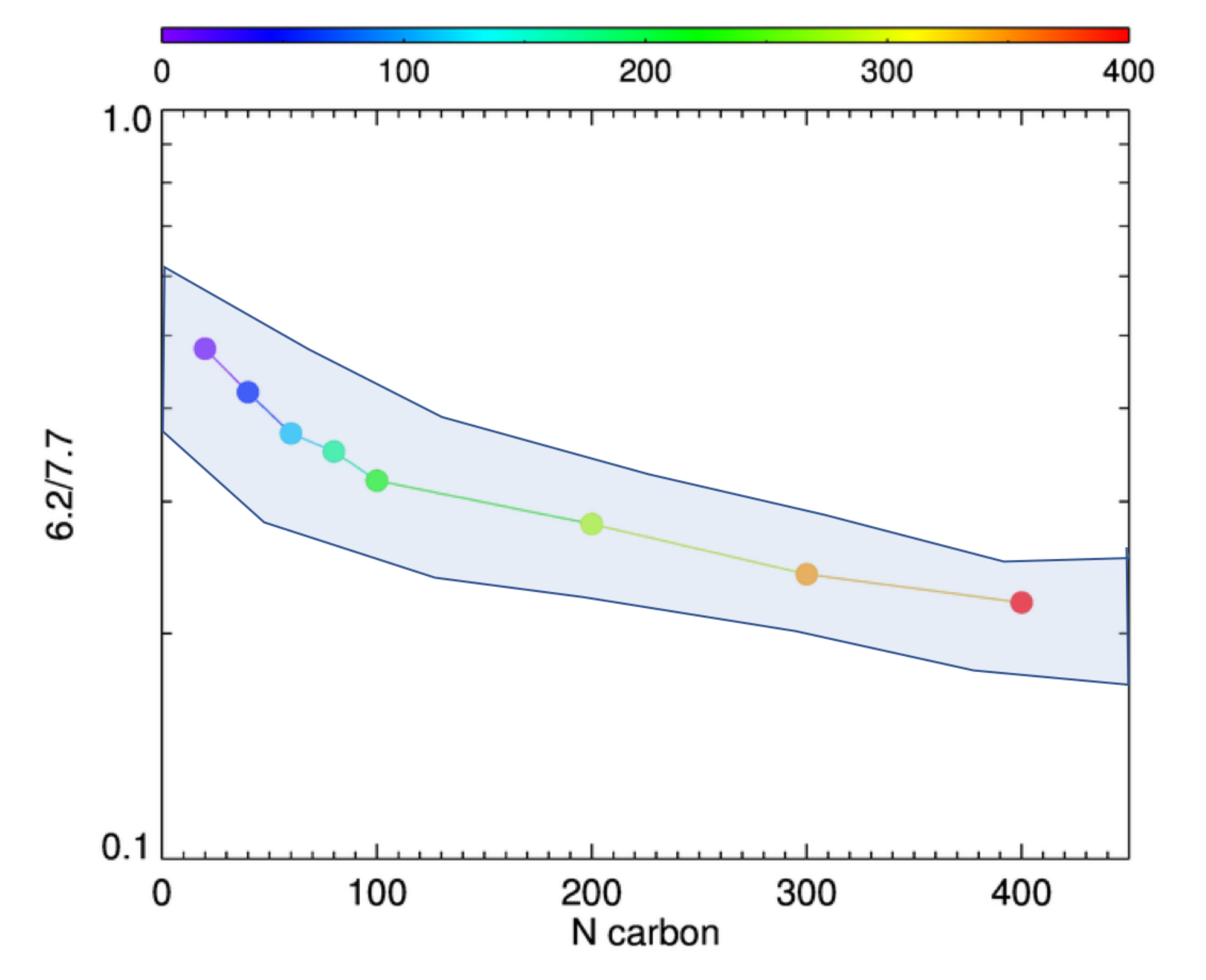}
    \caption{The 6.2$/$7.7 intensity ratio as a function of N$_{C}$
      for neutral PAHs exposed to the Galaxy ISRF radiation. The shaded
      region corresponds to the spread of values of each individual
      PAH considered in this study (see Section 3.) }
  \end{figure}

We find that the 6.2$/$7.7 ratio
tracks N$_{C}$ with the ratio decreasing by a factor of $\sim$2
between small and large values of N$_{C}$.
  We note that \cite{mara20} found significant scatter in the
  6.2$/$7.7 ratio as a function of N$_{C}$ and argued against the
  effectiveness of the ratio to trace PAH size. A number of reasons
  could be responsible for the different outcomes: First, there are
  differences in the measured PAH band intensities.  \cite{mara20}
  measure PAH bands by summing the flux between pre-determined
  wavelength ranges and apply no separate treatment for the 6-9 $\mu$m
  emission plateau. In the present work (outlined in Section 3), a set
  of Lorenztians are fit to the PAH emission features. Our methodology
  follows the profile decomposition method described by
  \citet{smith07} although in the case of the DFT spectra there is
  obviously no dust continuum emission. In this work each PAH feature
  is fit by a Lorentzian profile where the width and the central
  wavelength of the feature are allowed to vary as has been the case
  for fitting extragalactic PAH spectra from Spitzer and ISO (see
  \citealp{smith07} and references theirein).
    While the two approaches result in relative agreement in the
    fluxes measured for the ``isolated'' PAH features such as the 3.3 and
    the 11.3 $\mu$m, there are significant differences in the measured
    fluxes for the features in the 6-9 $\mu$m wavelength regime.
The DFT spectra often show significant (sub-)structure around the 5-9 $\mu$m
range with the complexity increasing with increasing number of carbons. As 
    e.g., \citealp{ricca12} point out the number of C–C and C–H
    bonds increases with increasing PAH size. This leads to an
    increase in the possible ways that the C-C and the C-H modes can
    couple which subsequently results in an increase in the number of
    features in the 6-9 $\mu$m wavelength regime.
  
\begin{figure}
\centering
\includegraphics[width=9cm]{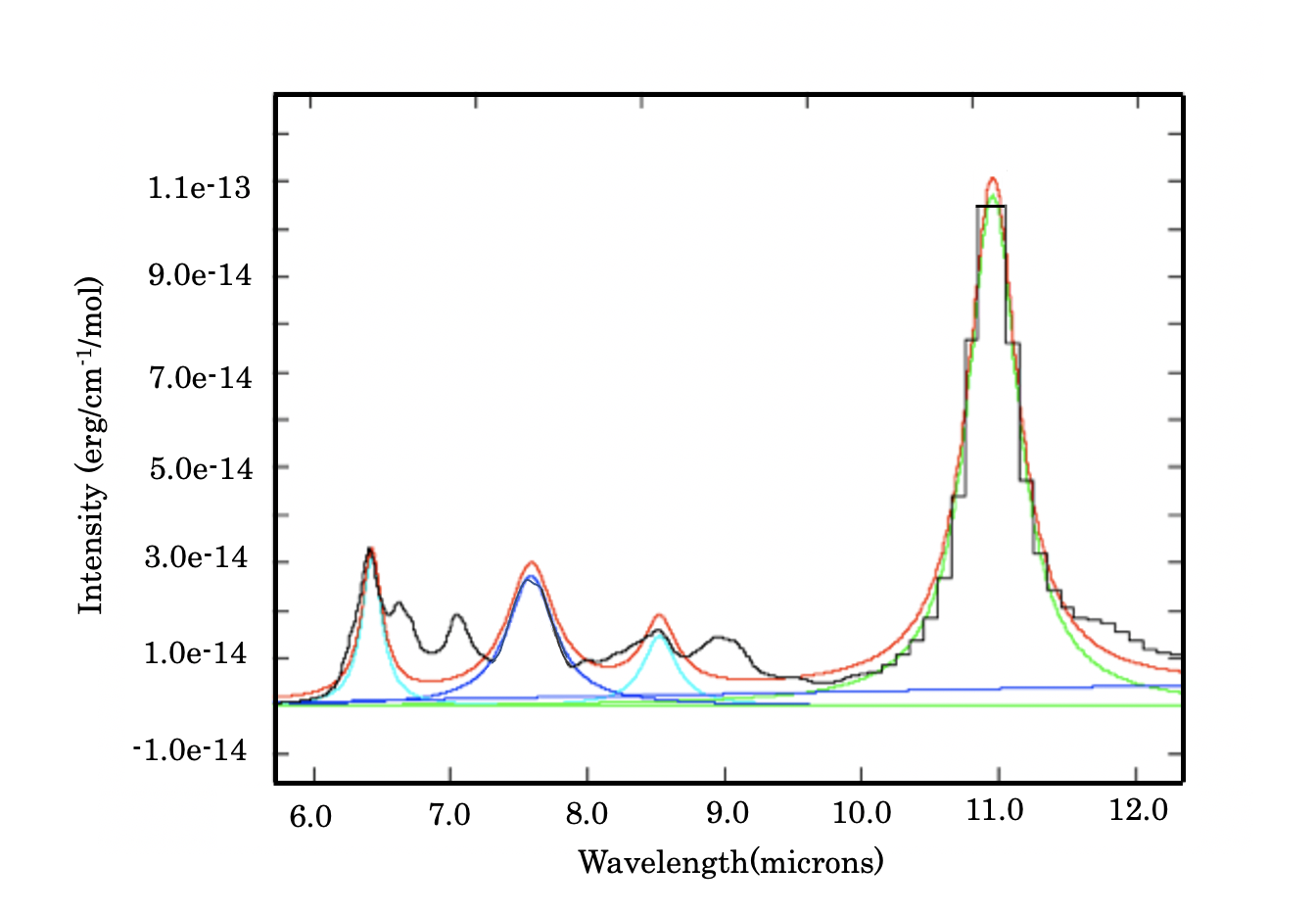}
\caption{ The 6-12 $\mu$m range of the
  averaged PAH spectrum in the 100$<$N$_{C}<$150 bin. The cyan,
  blue and green lines denote the Lorentzian fits to the 6.2, 7.7 and
  11.2 $\mu$m PAH features (see text for details of the fits).}
  \end{figure}

  The second, and perhaps, more important reason for the differences is
 the bin-averaging that we employ in our analysis. Given that the
 focus of the present work is on PAH spectra from extragalactic
 sources (and therefore beam averaged spectra of a number of individual PAHs)
 we have bin-averaged the DFT spectra based on bins of N$_{C}$ shown
 in Fig. 2. In addition, for the final averaged spectra in each bin we
 have excluded those spectra for which the flux
in a given PAH band was found to deviate by more than 3$\sigma$ from the average
value determined for that band in that specific bin. Figure 6 shows
the 6-12 $\mu$m range of the averaged spectrum in the
100$<$N$_{C}<$150 bin where each PAH feature is fit with a separate
Lorentzian centered at the nominal wavelength. Our fits for the 6.2
(cyan) and 7.7 (blue) PAH features result in a ratio of 6.2$/$7.7
$\sim$0.35 in agreement with the band ratio vs. N$_{c}$ plot shown in
Figure 5.

Next, we consider the 11.3$/$3.3 ratio as a function of N$_{C}$, shown 
in Fig. 7.
As already noted, both the 3.3 and 11.3 $\mu$m bands originate
from stretching and bending of the C-H bonds therefore their ratio could be suitable
to trace PAH molecular size.
A number of authors (e.g., \citealp{ricca12, mori12, croiset16,
mara20} have used the
ratio to trace PAH size distribution. We find that
the 11.3$/$3.3 ratio tracks N$_{C}$ with the ratio increasing by a
factor of $\sim$10 between small and large N$_{C}$ values. 
While the 11.3$/$3.3 can be used to trace PAH size distribution, 
it is worth pointing out two factors that
can influence the value of this ratio in astronomical sources:
the first, is the sensitivity of the 3.3 $\mu$m band
on the hardness of the radiation field. As is shown in Figure 1, the
strength of the 3.3 $\mu$m feature is very sensitive to changes in the
hardness of the radiation field it is exposed to. In addition, the
11.3$/$3.3 $\mu$m ratio is also affected by extinction as the two
features are bound to experience different levels of extinction
because of the wavelength difference. In a recent study, \citet{lai20}
investigated the effect of obscuration on PAH bands and found that
among all PAH bands, the 3.3 $\mu$m feature is the one that is most
susceptible to dust attenuation. In addition, \citet{ahc20}
investigated the effect of extinction on the 11.3 $\mu$m feature.
While extinction is bound to
affect all PAH band ratios, the 6.2$/$7.7 $\mu$m ratio is less
susceptible because of the proximity in the wavelengths of the two  
features. We discuss the dependence of the 11.3$/$3.3  ratio on the hardness of the
radiation field in detail in the next Section.

\begin{figure}
\centering
\includegraphics[width=9cm]{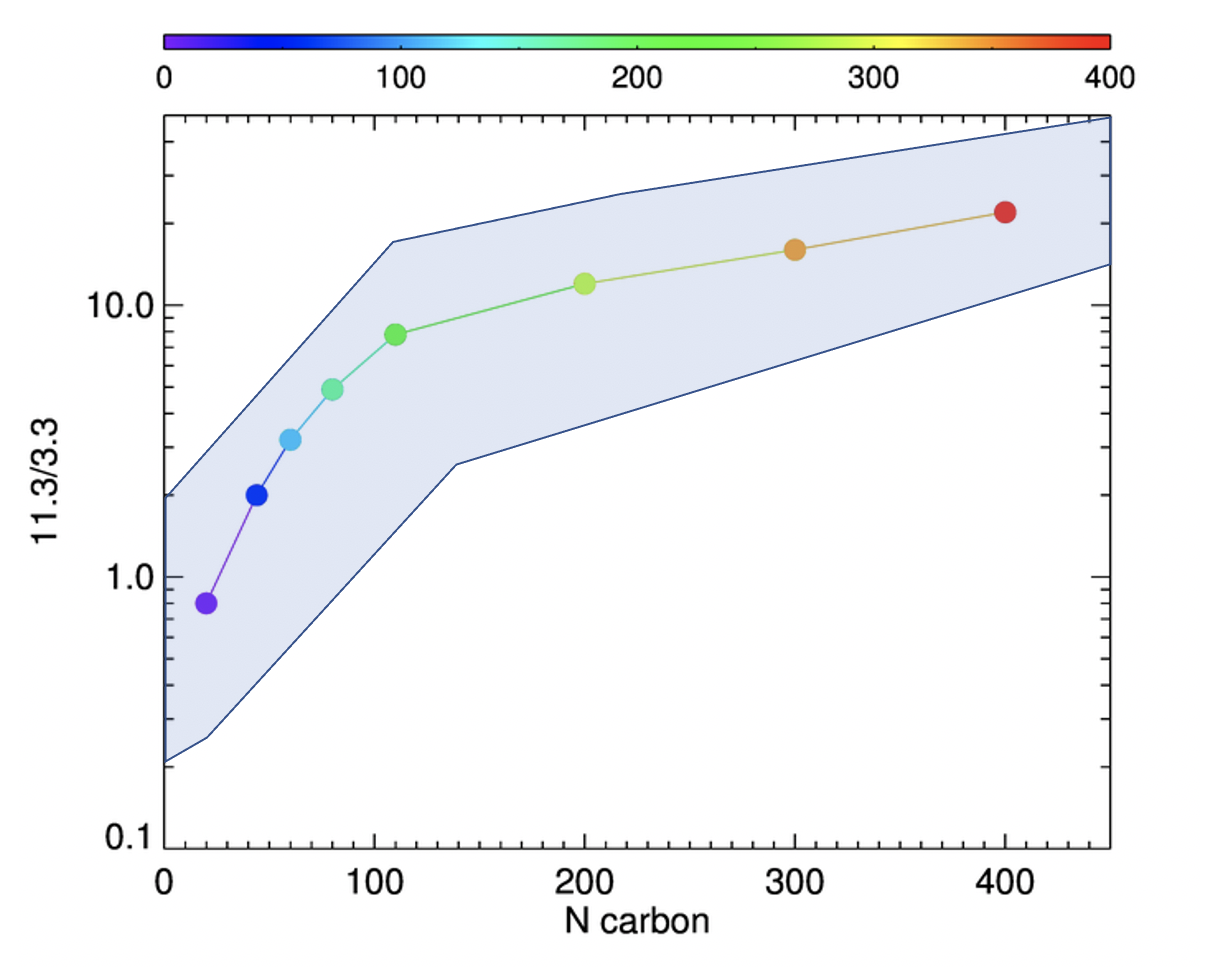}
    \caption{The 11.3$/$3.3 intensity ratio as a function of N$_{C}$
      for neutral PAHs exposed to the Galaxy ISRF radiation. The shaded
      region corresponds to the spread of values of each individual
      PAH considered in this study (see Section 3.)    }
  \end{figure}

\subsection{PAH band ratios vs hardness of the radiation field \& charge}

The fraction of ionised PAHs at any given time is determined by the balance between
photoionization and recombination with ambient electrons. The
photoionization is dependent upon the intensity of the radiation field
(expressed by G$_{0}$, the intensity of the radiation field in units
of the solar vicinity value 1.6$\times$10$^{6}$ W m$^{-2}$, \citet{habing68})
whereas the fraction of ambient electrons available for recombination
is parameterised by n$_{\rm e}$ the electron density.
Laboratory experiments and
theoretical studies \citep{defrees93, allam99,  elob01}
have shown that the ionization of PAHs enhances the
intensity of the band features in the 6 - 9 $\mu$m region relative to the
features in the 11 - 14 $\mu$m region. Therefore,
band ratios involving ionised to neutral bands (such as  6.2$/$11.3,
7.7$/$11.3)
should provide an estimate of the ionization of
PAHs. The PAH {\it ionization parameter} defined as
G$_{0}$T$^{1/2}_{gas}/$n$_{e}$, where T$_{gas}$ is the temperature of
the gas in K, can be linked to the fraction of ionized to neutral PAHs
and subsequently to PAH band strength ratios. In reality, both G$_{0}$
and the fraction of ionised PAHs also depend on the hardness of
the underlying radiation field that the PAHs are exposed to.

In order to robustly examine the
effect of ionization and of the hardness of the underlying radiation field
to PAH band ratios we consider two limiting cases one with low N$_{C}$
and one with large N$_{C}$
molecules. We use the `bin-averaged' values for PAHs at the low
(20$<$N$_{C}<$40) and
high end (200$<$N$_{C}<$400) of the PAH distribution,
as discussed in Section 3 and shown in Fig.2.
 In Fig. 8 we examine the
dependence of the 11.3$/$3.3 and 11.3$/$7.7 band ratios for {\it
  small} and {\it large} PAH
molecules on the hardness of the underlying field. The two band
ratios were measured when the molecules were exposed to radiation fields
of varying strength (from 6 to 12 eV). Both the 11.3$/$ 3.3 and
11.3$/$7.7 ratios are sensitive to the strength of the radiation field
they are exposed to, however, the 11.3$/$3.3 ratio shows a much
stronger dependence on the hardness of the underlying radiation
field. 
 \begin{figure*}
\centering
\includegraphics[width=16cm]{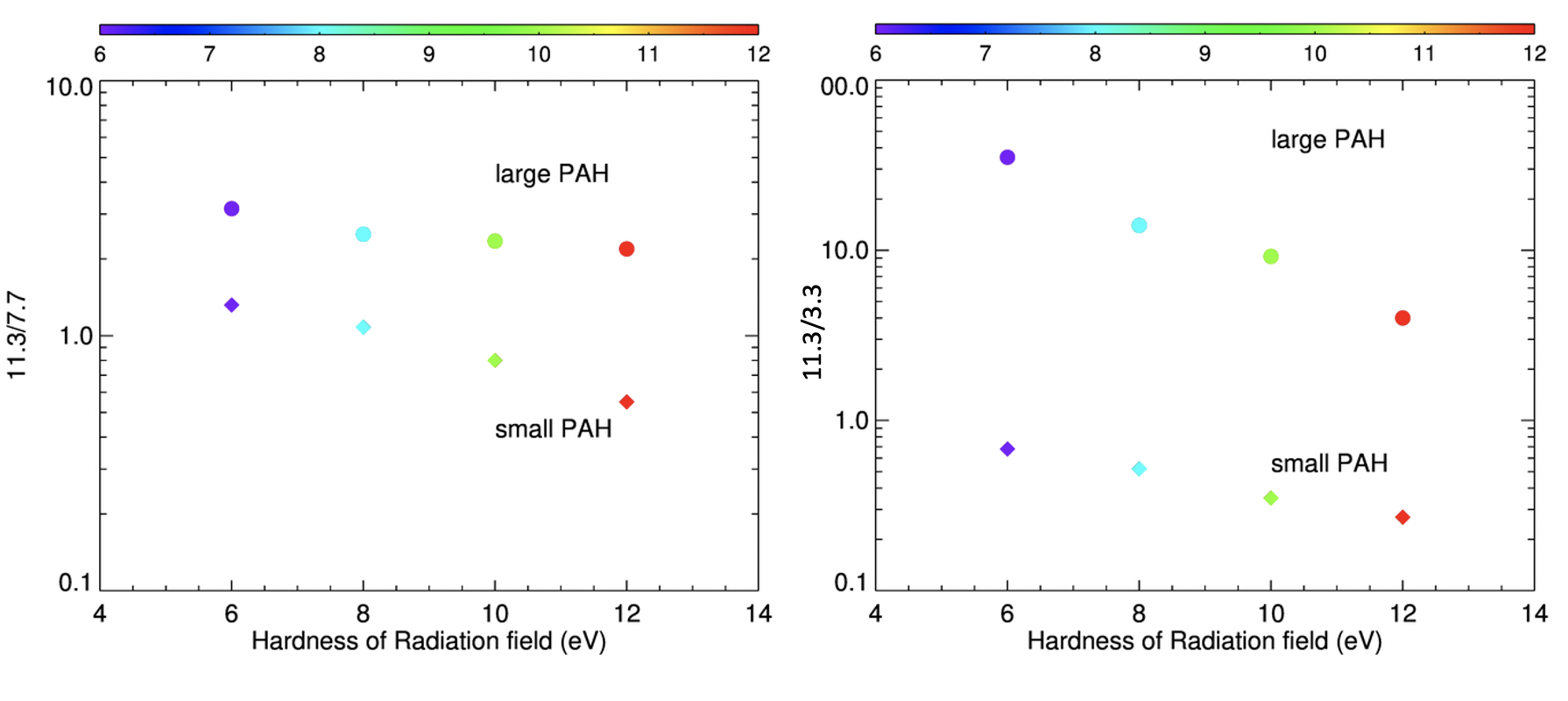}
    \caption{ The 11.3$/$7.7 (left) and 11.3$/$3.3 (right) intensity ratios
    calculated for the small bin-averaged molecule (bottom) and the largest bin-averaged
    (top) neutral PAH molecules as a function of  radiation fields with
    varying energy from 6 eV to 12 eV. The colour coding corresponds the
    hardness of the radiation field. }
  \end{figure*}
In Fig. 9 we plot the 11.3$/$3.3 and 11.3$/$7.7 band ratios for
 small and large PAH
molecules but this
time we vary
the fraction of neutral-to-ionized
molecules, starting from 100\% neutral to 100\% ionized. We find that
the 11.3$/$7.7 ratio shows a clear dependence on the ionization both
for small and large molecules. \citet{hony01} studied correlations
between PAH bands and mid-IR fine structure lines for a sample
of PDRs, HII regions and molecular clouds in the LMC.
They found that the ratio of 11.3$/$7.7 showed
only a weak dependence on the [NeII]$/$[NeIII] line ratio which they
used as an indicator of hardness for the incident radiation field. 
This result is in line with the finding from our models that indeed
the 11.3$/$7.7 ratio is a good tracer of the ionization fraction of
PAH molecules.
\citet{mori12} suggested that the 11.3$/$3.3 vs 7.7$/$11.3 ratio can be
used as a diagnostic tool for the radiation field conditions: a harder
radiation field will result in increased ionization fraction whereas a
softer field will decrease the number of ionized PAH molecules (their
Figures 6 \& 7). In agreement with \citet{mori12} we 
conclude that when combined these two band ratios could be used to probe the
physical conditions of the ionizing source. 
\begin{figure*}
\centering
\includegraphics[width=16cm]{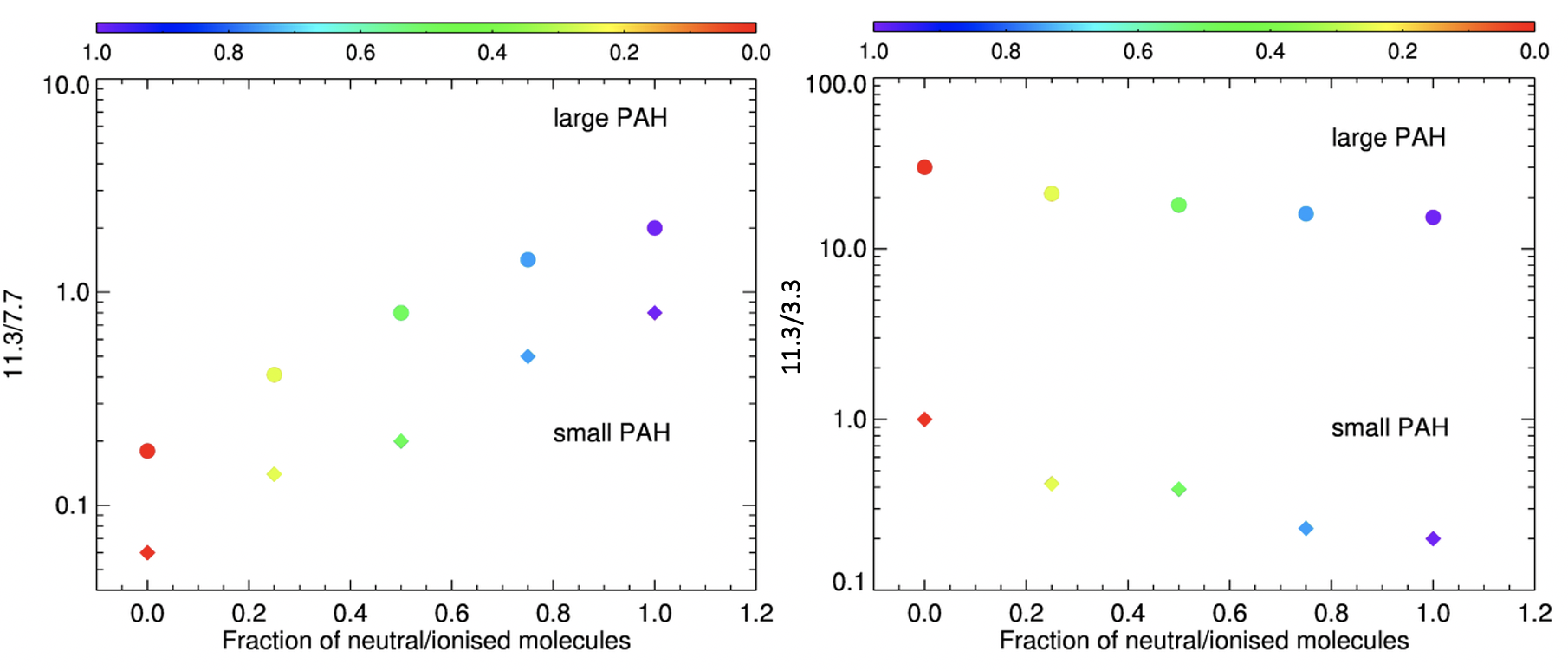}
    \caption{ The 11.3$/$7.7 (left) and the 11.3$/$3.3 (right)
      intensity ratios  calculated for the small bin-averaged (bottom) and 
      the large bin-averaged (top) PAH molecules with varying degrees of
      ionization starting from a neutral PAH (0\%) to a cation PAH
      (100\%).  The colour coding corresponds to fraction of neutral
      vs ionised molecules.}
  \end{figure*}

\subsection{The 6.2$/$7.7 vs 11.3$/$7.7 and 11.3$/$7.7 vs 11.3$/$3.3 ratios   }

In this section we bring together the results of the analysis described
  above as we seek to use the various PAH emission bands to
  construct a size-ionization-hardness grid.
The well established correlation between
6.2$/$7.7 and N$_{C}$, the steep dependence of the 11.3$/$7.7 on
the fraction of neutral vs ionised molecules and finally, the
ability of the 11.3$/$3.3 ratio to track the strength of the
underlying radiation field {\it define a three-dimensional plane which can
qualitatively} characterise the properties of PAHs and infer the
conditions of the ISM.
Figure 10 shows the 3D space defined by combining the three PAH band
ratios. The black tracks correspond to neutral PAHs while the magenta
dotted line represents the track for ionised PAHs. 
Figure 11 shows cuts 
along the 6.2$/$7.7 vs. 11.3$/$7.7 and 11.3$/$3.3 vs 11.3$/$7.7
planes and these are discussed in the next section.


\begin{figure*}
\centering
  \includegraphics[width=18cm]{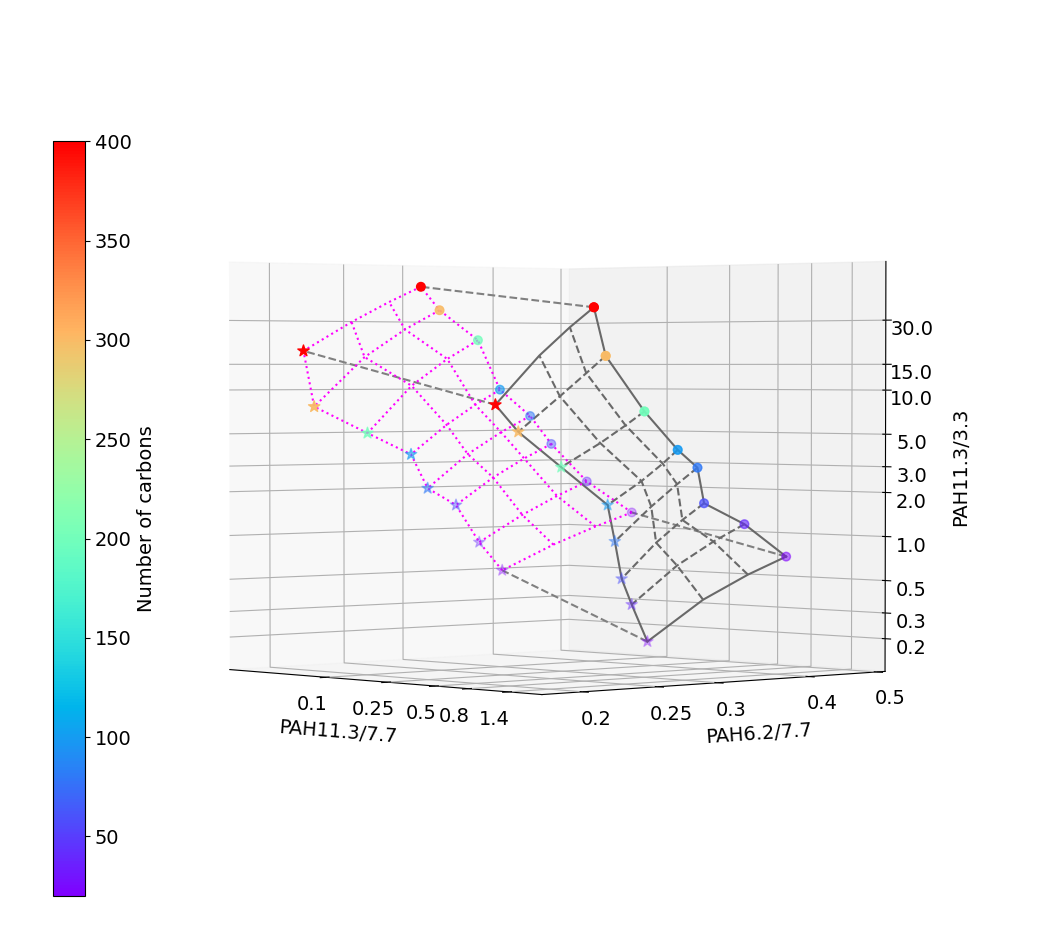}
  \caption{The 3D space defined by the 11.3$/$3.3 vs. 11.3$/$7.7
    vs. 6.2$/$7.7 ratios. The different colours correspond to
    different number of carbons. The filled circles corresponds to
    PAHs exposed  to ISRF while the filled stars corresponds
    to 1000 $\times$ISRF. The black tracks correspond to neutral PAHs
    while the magenta dotted lines corresponds to ionised PAHs.}

\end{figure*}

\begin{figure*}
\centering
\includegraphics[width=18cm]{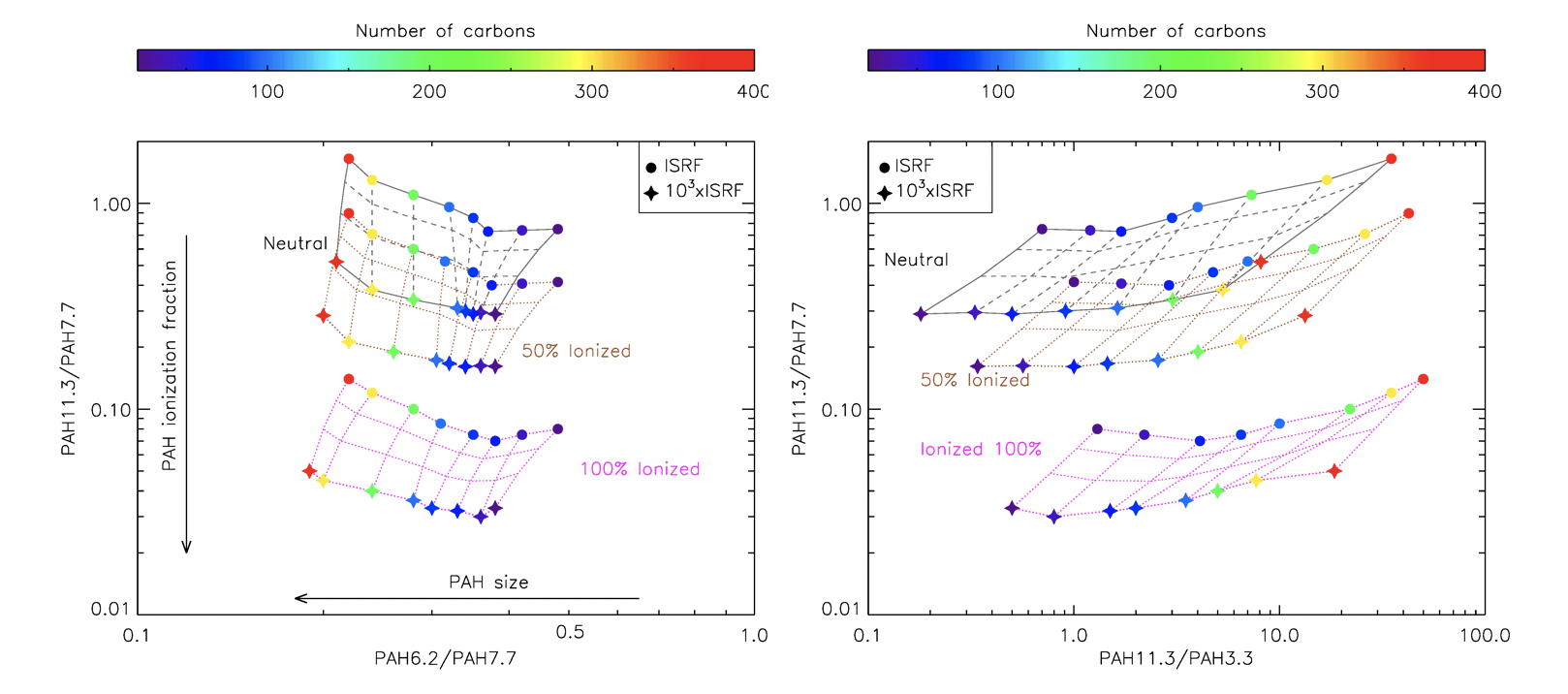}
  \caption{(Left:)The 6.2$/$7.7 vs. 11.3$/$7.7 ratio plot
    from DFT PAH spectra exposed to radiation fields
    with a range of energies from the ISRF (filled circles) to 10$^{3}
    \times$ISRF (stars). The top grid (black lines) represents neutral
    PAHs, the middle grid (brown lines) represents 50\% neutral and
    50\% ionised PAHs while the bottom one (magenta lines) corresponds
    to 100\% ionized PAHs.
    (Right:) The same grids for the 11.3$/$7.7 vs. 11.3$/$7.7 ratio plot
     from DFT PAH spectra exposed to radiation
      fields with a range of energies from ISRF to 10$^{3}\times$ISRF.  }
\end{figure*}



\section{Discussion: Comparison between Models and Data}

\subsection{The Sample Data}
The Infrared Spectrograph (IRS) on board Spitzer
and before then the Short Wavelength Spectrometer (SWS) onboard
ISO have acquired mid-IR spectra for a
large number of astronomical objects. Comprising four separate
modules, the IRS covered the entire 5.2 - 38 $\mu$m
in two resolution modes \citep{houck04}.
For the present study we focus on the strength of the 6.2, 7.7
and 11.3 PAH lines with values taken from published works on a
number of different samples of galaxies
including normal, Seyfert and dwarf galaxies. 
We only consider PAH band measurements from low
resolution spectroscopic observations. The line ratios are plotted
against the model grids generated from the synthetic spectra discussed earlier.
The galaxies considered here
come from the SINGS survey \citep{smith07}, the RSA sample of galaxies
\citep{ds10}
and blue compact dwarf galaxies \citep{hunt10}.
We stress here that we have not made any attempts to re-reduce the
spectra in a homogenous manner.  A more detailed analysis based on
IRS spectral maps will be presented in a future paper \citet{igb20}.

Fig. 12 shows the 6.2$/$7.7 vs. 11.3$/$7.7 band ratios for the
various galaxy samples outlined above. The tracks shown are for
neutral PAHs (black grid)
starting from small PAHs (N$_{c}$=20) to large PAHs (N$_{c}$=400)
and exposed to a radiation field ranging from ISRF to 10$^{3} \times$
ISRF as discussed in Section 3. In addition we show the track for
ionised PAHs (we only show the ISRF case for the ionised PAHs to
maintain a clear view of the location of the majority of the galaxies
on the plot). 

The majority of galaxies seem to preferentially lie at the
location of large PAHs with no noticeable difference between the
location of LINERs and AGN. {\hii} galaxies appear to favour
slightly smaller PAH molecules although with the current data it is
difficult to fully quantify the differences. 
Dwarf galaxies display a wide range of band ratios with roughly half
of the sample galaxies falling outside the range covered by the current grid.
Likewise, a small fraction of the LINERs also appear to fall outside
the grid. While a full discussion of why such sources may be located outside the grid
will appear in a forthcoming paper \citet{igb20} we
note here that it is likely that the harsh ISM conditions often
encountered in Dwarf galaxies may be responsible for the presence
of catacondensed PAH molecules (those exhibiting open and irregular structure).
In his recent review \citet{li20} suggests that emission spectra from
catacondensed molecules have a higher 11.3$/$7.7 ratio.
It is also likely that such extreme 11.3/7.7 ratios could be due to incomplete
subtraction of the continuum emission underlying the PAH
bands. The higher sensitivity and spectral resolution afforded with
the upcoming James Webb Space Telescope (JWST) is expected to shed
light on PAHs in a variety of extragalactic conditions.
It is, however, clear
that very few galaxies lie in the location of the grid where
20$<$N$_{C}<$50. 
In terms of ionization, it is evident that most galaxies require a
fraction of ionised PAHs between 25\% to 50\% with some
normal$/$starforming galaxies requiring $>$75\% ionised PAH 
molecules.

 \begin{figure*}
\centering
\includegraphics[width=18cm]{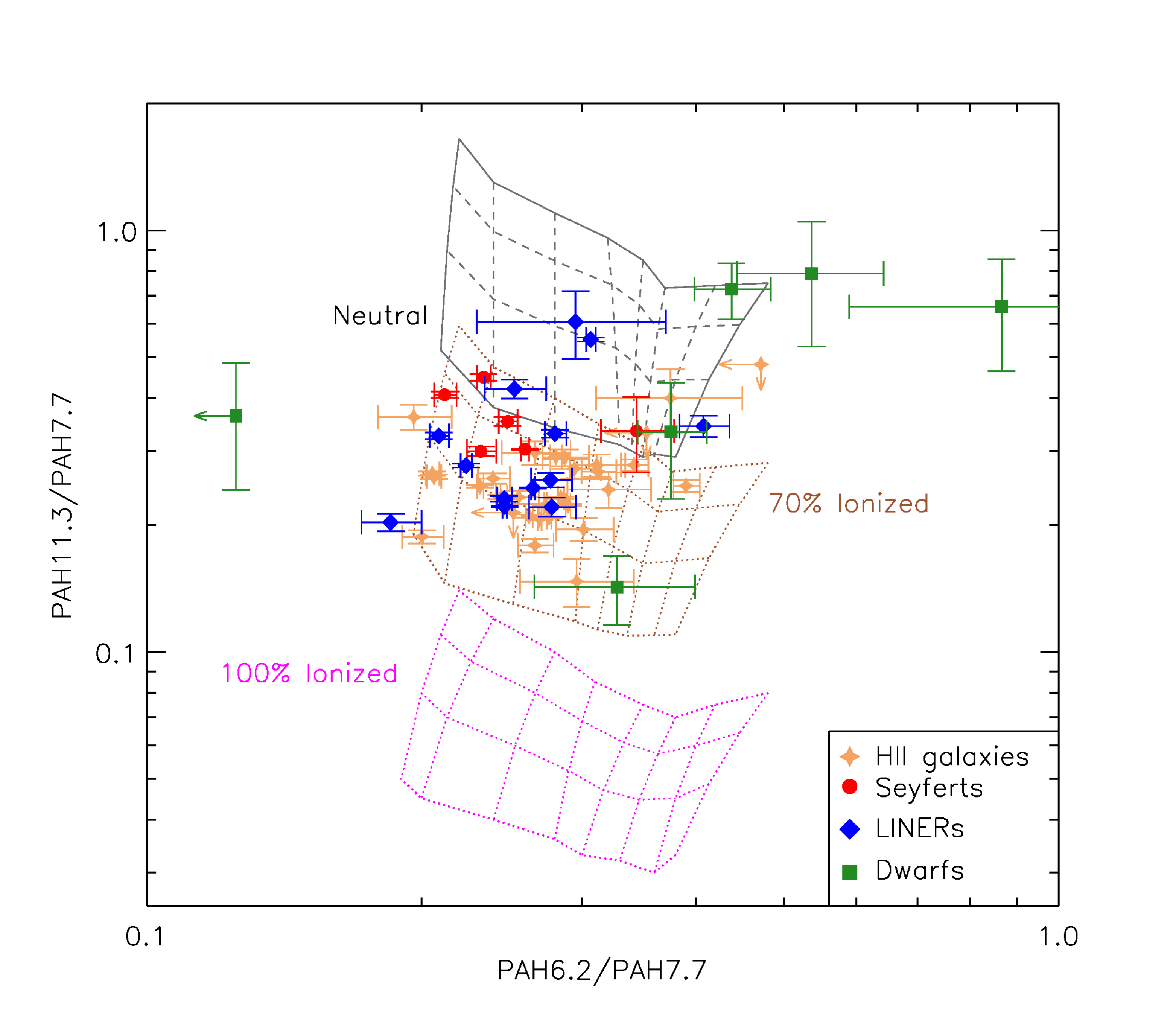}
    \caption{The size-ionization grid (same as in fig. 10) showing
      PAH band ratios for normal (HII) galaxies (orange stars), LINERS
      (blue diamonds),
      Seyferts (red circles) and dwarf galaxies (green squares). The
      top grid corresponds to neutral PAHs, the middle grid
      corresponds to a mixture of 70\% ionised - 100\% neutral and the
      bottom one corresponds to the fully ionised PAH grid.}
\end{figure*}

\section{Conclusions}
Based on DFT-computed PAH spectra we have presented a method to
investigate the size, charge and hardness of the radiation field based
on combinations of PAH band ratios. The DFT spectra were computed for
molecules with varying number of carbons in the range of 20$<$ N$_{C}<$400.
The spectra were generated using the Gaussian software
and
supplemented with
spectra from the NASA Ames PAH IR Spectroscopic Database. The spectra
were stacked in bins of increasing size and in each bin an `averaged'
spectrum was calculated. 3$\sigma$ outliers were removed from further
analysis.
The spectra were subsequently ``exposed'' to radiation fields
corresponding to multiples
of the ISRF, the interstellar radiation field of our Galaxy. We
considered neutral as well as cationic molecules. Our findings can be
summarised as follows:
\begin{enumerate}

\item{ PAH sizes: we have explored several band ratios to
determine the optimal
ones to trace the PAH molecular size. We investigated
 combinations of PAH band ratios originating from the bending or
 stretching of C-H or C-C
bonds and focus in particular on the 6.2$/$7.7 and$/$or 11.3$/$3.3 ratios.
Both ratios can be used to track PAH molecular size. The 6.2$/$7.7 ratio
changes by a factor of 2 between small and large molecules and  
is less sensitive to interstellar extinction since the wavelengths of the
features lie close together.
The 11.3$/$3.3 varies by a factor of 10 between small and large
molecules although the ratio is sensitive to extinction (see
discussion in \citet{lai20} for the 3.3 $\mu$m and \citet{ahc20} for
the 11.3 $\mu$m). In addition, this ratio shows a strong dependance
on the hardness of the radiation field.}

\item{Hardness of the underlying radiation field: we have explored,
 the influence of the hardness of the
    underlying radiation field on PAH band ratios. We explored the
    effects of varying the intensity of the radiation field on two
    limiting cases, the bin-averaged small and the bin-averaged
    large PAH
    molecules. We found that the 11.3$/$3.3 band ratio shows a 
   good  correlation with the intensity of the radiation field with the
   ratio decreasing with increasing energy. A similar trend is noted
   by \citet{mori12} for their Group A sources.}

\item{Fraction of neutral-to-ionized PAH molecules: we explored
  ratios sensitive to the fraction of neutral-to-ionized
  molecules. Amongst the band ratios examined we found that the
  11.3$/$7.7 increases steeply as we progress from 100\% neutral to
  100\% ionized molecules. }

\item{We have compared published Spitzer IRS data from samples of
    normal, star-forming, dwarf galaxies as well as active galactic
    nuclei to our grids. We find that the majority of galaxies tend to
  favour large PAHs exposed to radiation fields of moderate
  strength. The majority of the sample galaxies favour an ionization
  (fraction of neutral to ionised molecules) of $<$1 with the
  star-forming galaxies requiring a higher fraction of charged
  PAHs. It is obvious from Fig. 12 that we need to
  investigate the behaviour of large molecules with N$_{C} >500$ as
  the majority of the band ratios seen in galaxies tend to favour
  large molecules. This issue will be addressed in a forthcoming paper
(Kerkeni et al., in prep.)}

\item{We have demonstrated that by combining the three ratios
    6.2$/$7.7, 11.3$/$3.3 and
    11.3$/$7.7, we can qualitatively constraint the size and the physical conditions
    in the regions where PAHs originate. The upcoming availability of
    JWST will enable measurments of these bands in a number of
    galactic and extragalactic sources. In addition, we will be able
    to study these ratios within regions of galaxies and map out
    the influence of the underlying radiation source (star or AGN) on the
    immediate surroundings and explore the fate of small$/$large
    PAHs. }

 \end{enumerate}

\section*{Acknowledgements}
We thank the anonymous referee for his$/$her comments.
DR and IGB acknowledge support from STFC through grant
ST$/$S000488$/$1. DR and BK acknowledge support from the University of
Oxford John Fell Fund.  AAH acknowledges support from grant
PGC2018-094671-B-I00 (MCIU/AEI/FEDER,UE). AAH and MPS work was done
under project No. MDM-2017-0737 Unidad de Excelencia ``Mar\'{\i}a de
Maeztu''- Centro de Astrobiolog\'{\i}a (INTA-CSIC).
The authors would like to acknowledge the use of the University of
Oxford Advanced Research Computing (ARC) facility in carrying out this work.

\section*{Data Availability}
The data underlying this article will be shared on reasonable request
to the corresponding author.








\appendix

\section{PAH Molecules}

Table A1 lists all the molecules used in this study. The numbers reported
correspond to the UIDs in the NASA Ames Database for neutral and
cationic molecules. Where there is no UID entry, the frequencies of
those molecules have been computed  by us. A discussion of the
methodology followed to compute the vibrational spectra of various PAH molecules
will appear in Kerkeni et al in prep. as described in Section 2. 
 \begin{table*}
  \caption{Molecules used in this study}
  \label{tab:landscape}
  \begin{tabular}{cccccc}
    \hline
Molecule&Neutral&Cation&Molecule&Neutral&Cation\\
    \hline
C22H12&2347 &3156&C22H14 &301,305,307&302,306,308 \\
C24H12$^{1}$& &&C27H13&66 &65\\
C30H14$^{1}$ & & &C31H15&3226 & 3227 \\
    C32H14$^{1}$& &&C34H16&3161 &3162\\
    C35H15&3229 &3230&C36H16$^{1}$& & \\
    C37H15& 3232&3234&C40H18&131, 533, 625&132,540,626\\
  C42H18$^{1}$& && C42H22&137,156&138,158\\
    C43H17&3235&3236&C44H18$^{1}$&&\\
    C44H20&140&141&C45H15&718&722\\
    C45H17&3238&3239&C46H18&3169&3170\\
    C47H17&76&77 &C48H18$^{1}$& &\\
C48H20&100,146& 101,147&C48H22&143, 3941, 3942,3943 & 144,3957,3958, 3959\\
    C51H19&3241 &3242&C52H18&3173&3174\\
    C54H18$^{1}$&&&C54H20&3176&3177\\
C55H19&3244&3245 &C56H20&3179&3180 \\
C57H19&646&644 &C58H20$^{1}$& & \\
C59H21&3247 &3248&C62H20$^{1}$& & \\
    C63H21&727 &730&C64H20&3188&3189\\
    C64H22&3191 & 3192&C65H20$^{1}$& & \\
    C66H18& 713,714& 715, 716&C66H20$^{1}$& &\\
    C67H21&656 &654&C67H23&3252, 3255 &3253, 3256\\
    C71H21&659 & 657& C72H22& 3194& 3195 \\
    C73H21&3257 &32578&C76H22&3197 &3198\\
    C77H23&3260 &3261&C78H22$^{1}$& & \\
    C82H24&561,3203 & 562,3204&C83H23&3265 &3266\\
 C85H23&3268 & 3269 &C87H23& 649&647 \\
    C88H24$^{1}$& & &C90H24&638 &639\\
    C91H25&3273&3274 &C94H24& 3211&3212\\
    C95H25&3277 &3278&    C95H27&3280 &3281\\
 C96H23&693, 696 & 694,697 &C96H24$^{1}$& & \\
    C98H28&565,567&566,568&C102H26&177,180 3214&178,181,3215\\
    C103H27& 3285&3286 &C107H27&3287,3289 &3288, 3290\\
    C108H26&3217 & 3218&C110H26&162 &163\\
    C111H27&3291 & 3292&C112H26$^{1}$\\
                     C115H29& 3296&3297&C128H28&631 & 632\\
    C130H28&168 &169&C138H30&772 &773\\
    C142H30&760,763,766 &761,764,767&C144H30&756,641 &757, 642\\
C146H30&746,749,752 &747, 750,753&C148H30&743 &744\\
    C150H30&612 &613&C170H32&619 &623\\
    C190H34&634&738&C210H36&742&740\\
    C294H42&616& & C384H48$^{1}$&617& \\
    \hline
\multicolumn{6}{c}{$^1$ vibrational frequencies for these molecules have been
    computed independently.}
\end{tabular}
 \end{table*}

\section{Ionized PAHs}


In addition to the neutral PAHs we have followed the same procedure to
compute DFT-spectra for PAH cations (corresponding to the neutral PAHs
used in our study). To compute the spectra of the cations we followed
the same steps as outlined in Section 2, using the same B3LYP
functionsal aling with thr 4-31G basis set. The DFT- computed
transition frequencies for the PAH cations 
were subsequently convolved with a specific band shape, line width and
emission temperature to convert them to emission spectra.
Applying the same methodology as with the neutral PAH spectra,
we placed the cation spectra in bins (see Fig. 2) and  for each bin we
computed an average spectrum by determining the arithmetic mean and
dividing the average value
by the number of molecules that were assigned to that bin. Fig. A1
shows the relative intensities of the 3.3, 6.2, 7.7, 8.6, 11.3 $\mu$m
cation PAH emission
bands calculated based on the ISRF emission model. The Figure
shows the average intensities as well as the 3$\sigma$ spread around this
value for every bin.
\begin{figure*}
 \centering
 \includegraphics[width=17cm, angle=0]{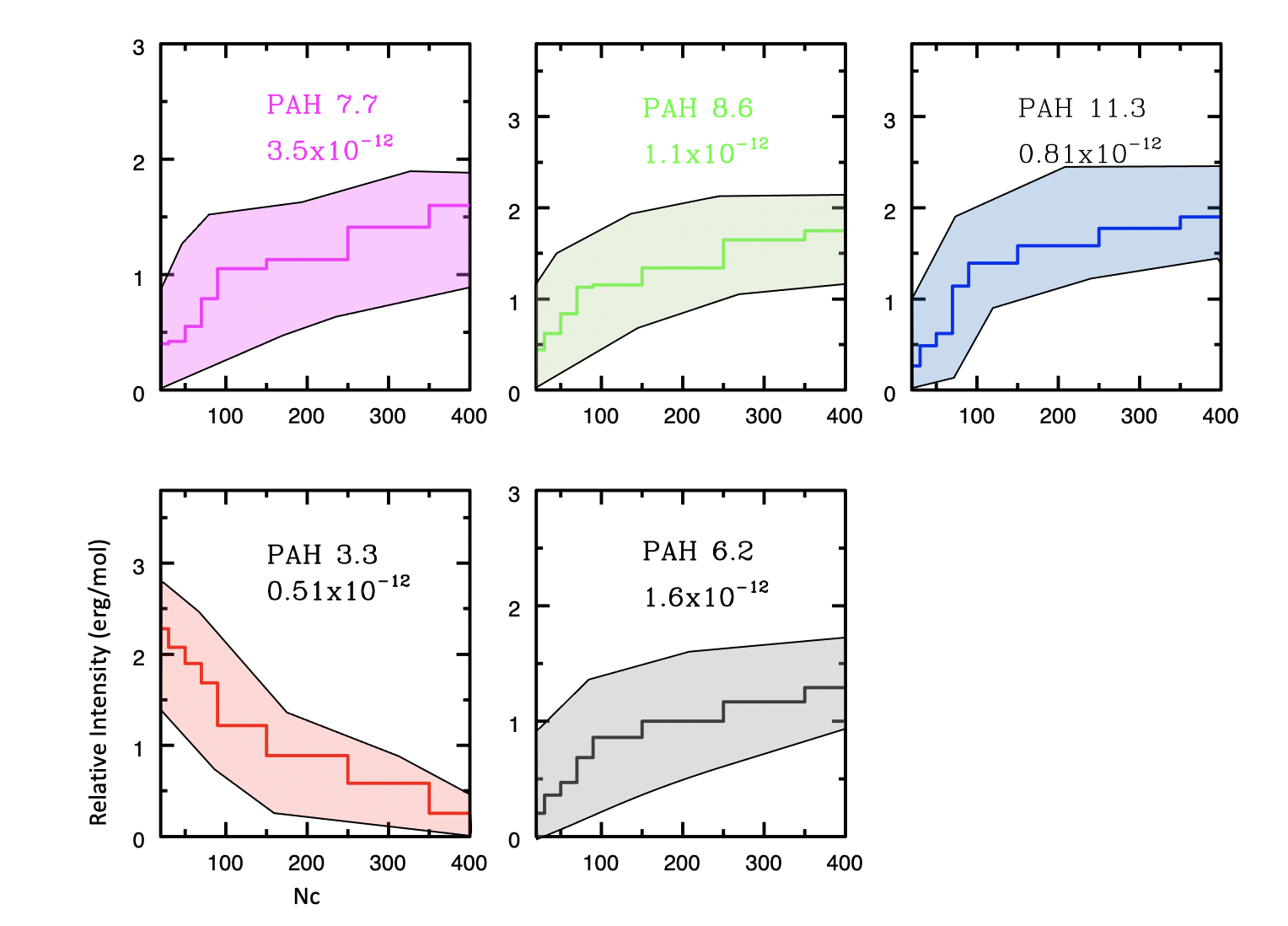}
    \caption{ Intensities for cationic PAHs (3.3, 6.2, 7.7, 8.6 and
      11.3 $\mu$m) as a function  of N$_{\rm C}$, with normalisations
      shown in each panel.
      The average value in each bin
     is denoted with the line while the shaded regions denote
     the  spread of
     values in each bin. PAHs with  intensity values greater than 3$\sigma$ have
     been excluded. The plots are normalised to N
     The intensities have been calculated for the ISRF
      radiation field.  Colour coding:
      3.3$\mu$m (red), 6.2 $\mu$m (grey), 7.7$\mu$m (magenta), 8.6 $\mu$m
      (green), 11.3 $\mu$m (blue). }
  \end{figure*}


\bsp	
\label{lastpage}
\end{document}